\documentclass[Afour,sagev,times]{sagexj}

\usepackage{moreverb,url}

\usepackage[colorlinks,bookmarksopen,bookmarksnumbered,citecolor=red,urlcolor=red]{hyperref}

\newcommand\BibTeX{{\rmfamily B\kern-.05em \textsc{i\kern-.025em b}\kern-.08em
T\kern-.1667em\lower.7ex\hbox{E}\kern-.125emX}}

\def\volumeyear{2020}

\begin{document}

\runninghead{Faddoul et al.}

\title{A longitudinal analysis of YouTube's promotion of conspiracy videos}

\author{Marc Faddoul\affilnum{1}, Guillaume Chaslot\affilnum{3}, and Hany Farid\affilnum{1,2}}

\affiliation{
\affilnum{1}School of Information, University of California, Berkeley\\
\affilnum{2}Electrical Engineering \& Computer Sciences, University of California, Berkeley\\
\affilnum{3}Mozilla Foundation}

\corrauth{Hany Farid,
School of Information,
UC Berkeley}
\email{hfarid@berkeley.edu}

\begin{abstract}
Conspiracy theories have flourished on social media, raising concerns that such content is fueling the spread of disinformation, supporting extremist ideologies, and in some cases, leading to violence. Under increased scrutiny and pressure from legislators and the public, YouTube announced efforts to change their recommendation algorithms so that the most egregious conspiracy videos are demoted and demonetized. To verify this claim, we have developed a classifier for automatically determining if a video is conspiratorial (e.g., the moon landing was faked, the pyramids of Giza were built by aliens, end of the world prophecies, etc.). We coupled this classifier with an emulation of YouTube's watch-next algorithm on more than a thousand popular informational channels to obtain a year-long picture of the videos actively promoted by YouTube. We also obtained trends of the so-called filter-bubble effect for conspiracy theories.
\end{abstract}

\maketitle

\section*{Introduction}
\label{sec:introduction}

By allowing for a wide range of opinions to coexist, social media has allowed for an open exchange of ideas. There have, however, been concerns that the recommendation engines which power these services amplify sensational content because of its tendency to generate more engagement. The algorithmic promotion of conspiracy theories by YouTube's recommendation engine, in particular, has recently been of growing concern to academics~\cite{lewis_2018, bessi_2016, song_2017, greene_2014, samory_2018, nyt_tufecki, climate}, legislators~\cite{guardian-senator}, and the public~\cite{mozilla, guardian, buzzfeed, nyt_roose, wired, avaaz}. In August 2019, the FBI introduced fringe conspiracy theories as a domestic terrorist threat, due to the increasing number of violent incidents motivated by such beliefs~\cite{fbi_report}.
 
Some $70\%$ of watched content on YouTube is recommended content~\cite{cnet}, in which YouTube algorithms promote videos based on a number of factors including optimizing for user-engagement or view-time. Because conspiracy theories generally feature novel and provoking content, they tend to yield higher that average engagement~\cite{hussain_2018}. The recommendation algorithms are thus vulnerable to sparking a reinforcing feedback loop~\cite{Zhao2019} in which more conspiracy theories are recommended and consumed~\cite{chaslot_2017}.

YouTube has, however, contested this narrative with three main counter-arguments~\cite{mohan}: (1) According to YouTube's Chief Product Officer Neal Mohan, "it is not the case that “extreme” content drives a higher version of engagement"; (2) The company claims that view-time is not the only metric accounted for by the recommendation algorithm; and (3) Recommendations are made within a spectrum of opinions, leaving users the option to engage or not with specific content. 

We are skeptical that these counter-arguments are consistent with what we and others qualitatively have been seeing play out on YouTube for the past several years. In particular: (1) according to Facebook’s CEO Mark Zuckerberg, extreme content does drive more engagement on social media~\cite{zuckerberg_2018}; (2) Although view-time might not be the only metric driving the recommendation algorithms, YouTube has not fully explained what the other factors are, or their relative contributions. It is unarguable, nevertheless, that keeping users engaged remains the main driver for YouTube’s advertising revenues~\cite{blog_0, youtube_deepNN}; and (3) While recommendations may span a spectrum, users preferably engage with content that conforms to their existing world view~\cite{confirmation_bias}.

Nonetheless, in January of 2019 YouTube announced efforts to reduce "recommendations of borderline content and content that could misinform users in harmful ways -- such as videos promoting a phony miracle cure for a serious illness, claiming the earth is flat, or making blatantly false claims about historic events like 9/11"~\cite{blog_1}. This effort complemented a previous initiative to include direct links to Wikipedia with videos related to conspiratorial topics.~\cite{wired_wiki} In June of 2019, YouTube announced that their efforts led to a reduction of view-time from these recommendations by over $50\%$~\cite{blog_2}. In December of 2019, YouTube updated this estimate to $70\%$~\cite{blog_3}. Our analysis aims to better understand the nature and extent of YouTube's promotion of conspiratorial content.



\section*{Materials \& Methods}
\label{sec:methods}

\subsection{Recommendations}

YouTube makes algorithmic recommendations in several different places. We focus on the \textit{watch-next} algorithm, which is the system that recommends a video to be shown next when auto-play is enabled. YouTube distinguishes between two types of recommendations: \textit{recommended-for-you} videos are computed based on the user's previous viewing history and \textit{recommended} are not individualized. Our requests are made with a U.S.-based IP addresses, without any identifying cookie. There are, therefore, no \textit{recommended-for-you} videos.

Our method to emulate the recommendation engine is a two step process: we start by gathering a list of seed channels, and then generate recommendations starting from the videos posted by these channels.

The list of seed channels is obtained with a snowball method. We start with an initial list of $250$ of the most subscribed English YouTube channels. The last video posted by each of these seed channels is retrieved and the next $20$ watch-next recommendations are extracted. The channels associated with these recommendations are ranked by number of occurrences. The channel that has the largest number of recommendations, and is not part of the initial seed set, is added to the set of seed channels. This process is repeated until $12,000$ channels are gathered. 

To focus our computational resources on the parts of YouTube that are relevant to information and disinformation, we perform a cluster analysis~\cite{blondel2008fast} on these $12,000$ channels. We retain a single cluster of $1103$ channels which corresponds to news and information channels (e.g., BBC, CNN, FOX...). Since the unsupervised clustering is not perfect, we manually added $43$ channels that we considered to be consistent with the other information channels. This yielded a final list of $1146$ seed channels, then reduced to $1080$ by the end of the analysis after some channels were deleted or stalled.

\footnote{\label{data_url} The list of seed channels and the training set are available at \mbox{\url{https://github.com/youtube-dataset/conspiracy}}}

We then gathered the $20$ first recommendations from the watch-next algorithm starting from the last video uploaded by each of the seed channels everyday from October 2018 to February 2020. The top $1000$ most recommended videos on a given day were retained and used in our analysis. As described below, these videos were analyzed to determine which were predicted to be conspiratorial in nature.

%

\begin{table*}
\small
\begin{center}
\resizebox{\textwidth}{!}{%
\begin{tabular}{| p{1.3cm} | p{4.3cm} | p{5.2cm} | p{5.0cm}|}
\toprule
\textbf{\hspace{1cm}} & \textbf{Comments} & \textbf{Snippets} & \textbf{Transcripts} \\ 
\midrule
\textbf{Positive} &
 \emph{illuminati, evil, told, research, deep, hoax, global, control, killed, believe, autism,  satanic, they, aliens, info} & 
 \emph{conspiracy, warming, qanon, truth, hoax, prophecy, illuminati, supernatural, report,  jfk, deception, ufo, evidence, energy, mystery} & 
 \emph{information, all, nasa, weather, nothing, footage, see, warming, evidence, know, climate, vaccines, ancient, look, aluminum}  \\ 
 \hline
\textbf{Negative} &
\emph{cute, universe, eat, future, dog, left, content, game, cool, imagine, food, better, loved, quality, pay} & 
\emph{biggest, policy, big, sea, camera, sermon, party, round, november, live, hot, process, model, culture, duty} & 
\emph{gonna, really, like, sea, young, side, him, black, live, early, policy, think, away, agents, thank}  \\
\bottomrule
\end{tabular}}
\caption{Most discriminating words in the training set for positive (conspiratorial) and negative labels, ranked by TFIDF.}
\label{tab:words}
\end{center}
\end{table*}


\subsection{Training Set}
\label{dataset}

We collected a training set of conspiracy videos in an iterative process. An initial set of $200$ videos was collected from a book referencing top conspiracy theories on YouTube~\cite{201conspiracy}, and a set of videos harvested on 4chan and on the sub-reddits r/conspiracy, r/conspiracyhub, and r/HealthConspiracy. A comparable set of $200$ non-conspiratorial videos was collected by randomly scraping YouTube videos. These videos were manually curated to remove any potentially conspiratorial videos. As we began our analysis, we augmented these initial videos by adding any obviously mis-classified videos into the appropriate conspiratorial or non-conspiratorial training set, yielding a final set of $542$ conspiratorial videos and $568$ non-conspiratorial videos.

We are sensitive to the fact that classifying a video as conspiratorial is not always clear-cut. We endeavored to limit our training set to videos whose underlying thesis, by and large, satisfies the following criteria: (1) Explains events as secret plots by powerful forces rather than as overt activities or accidents; (2) Holds a view of the world that goes against scientific consensus; (3) Is not backed by evidence, but instead by information that was claimed to be obtained through privileged access; (4) Is self-filing or unfalsifiable.

\subsection{Text Classification}

A key component of our video classifier is \texttt{fastText}, a text-based classifier~\cite{fasttext}. This classifier takes a text sample as input, and predicts the probability that the sample belongs to a given class (e.g., a conspiratorial video).

The classifier begins by parsing the training data to define a vocabulary. Input text samples are then represented by a concatenation of a bag-of-words and bag of $n$-grams, as defined by the vocabulary. An embedding matrix projects this representation into a lower-dimensional space, after which a linear classifier is used to classify the text into one of two (or more) classes.

\subsection{Video Classification}

Our video classifier analyzes various text-based components of a video using individual classifier modules for each. These modules, described next, are followed by a second layer that combines their outputs to yield a final conspiracy likelihood.

\begin{enumerate}
    \item \textbf{The transcript of the video}, also called subtitles, can be uploaded by the creator or auto-generated by YouTube, and captures the content of the video. The transcript is scored by a \texttt{fastText} classifier. 
    \item \textbf{The video snippet} is the concatenation of the title, the description, and the tags of the video. The snippet renders the language used by the content creator to describe their video. The snippet is also scored by a \texttt{fastText} classifier. 
    \item \textbf{The content of the $200$ top comments} defined by YouTube's relevance metric (without replies). Each comment is individually scored by a \texttt{fastText} classifier. The score of a video is the median score of all its comments.
    \item \textbf{The perceived impact of the comments.} We use Google's Perspective API~\cite{perspective} to score each comment on the following properties: (1) toxicity; (2) spam; (3) unsubstantial; (4) threat; (5) incoherent; (6) profanity; and (7) inflammatory. This set of seven perspective scores for each comment is converted into a $35$-D feature vector for the whole video by taking the median value and standard deviation of each property ($14$ features) as well as the median value of the pair-wise products of each property ($21$ features). A logistic regression classifier is trained to predict the conspiracy likelihood of the video from this $35$-D feature vector.
\end{enumerate}

The output of these four modules is then fed into a final logistic regression layer to yield a prediction for the entire video.

The two layers of the pipeline are trained on distinct videos with a $100$-fold cross validation. Specifically, our training set of $1095$ videos is randomly split into a $60/40$ split. The $60\%$ is used to train the four modules of the first layer. The remaining $40\%$ of videos are scored by these four classification modules. These scores are then standardized into four feature vectors each with a mean of zero and unit variance. The zero-mean ensures that missing attributes have a null contribution (e.g.,~transcripts can be unavailable), while the unit variance allows us to compare the relative importance of each attribute in the model. The final logistic regression is then trained on the $40\%$ split to predict if a video is conspiratorial. We repeat this process with $100$ different $60/40$ splits. By averaging the $100$ logistic regression models, we obtain the final regression coefficients. Their relative weights are $52\%$ for the comments, $22\%$ for the snippet, $14\%$ for the caption and $12\%$ for the perspective score. 
\subsection{Model Accuracy}

\begin{figure}[t]
    \centering
    \includegraphics[width=0.9\columnwidth]{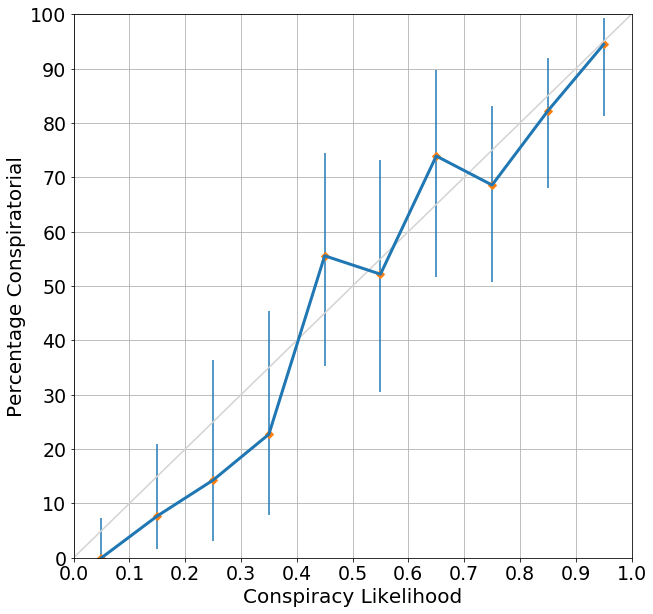}
    \caption{Percentage of videos labeled as conspiratorial by a human annotator plotted as a function of the estimated conspiracy likelihood, on a total of $340$ out-of-sample videos. The error bars correspond to Clopper-Pearson $95\%$ confidence intervals based on Beta distribution.}
    \label{fig:precision}
\end{figure}

To test the accuracy of our model, we manually classified $340$ videos not used in the training set. These videos were randomly sampled so that their score distribution is uniform between $0$ and $1$. Shown in Fig.~\ref{fig:precision} is the correlation between the conspiracy likelihood of the classifier (horizontal axis) and the percentage of videos rated as conspiratorial by a human annotator (vertical axis). With small fluctuations, our predicted conspiracy likelihood accurately predicts the actual likelihood of a video being conspiratorial, for example, $70\%$ of videos with a likelihood score of $0.7$ will be conspiratorial. With a threshold at $0.5$, the conspiracy classifier has a precision of $78\%$ and a recall of $86\%$.

From a more qualitative perspective, Table~\ref{tab:words} shows the words that are most statistically relevant to discriminating between conspiratorial and non-conspiratorial videos, as determined by \textit{term frequency inverse document frequency} (TFIDF)~\cite{tfidf}. Words that identify conspiracies seem reasonably diagnostic: they are either specific to a topic (e.g., \emph{aliens}, ~\emph{deep} - for Deep State, \emph{autism} - for vaccines), generic to conspiratorial narratives (e.g., \emph{deception}, \emph{control}) as well as, ironically, words that characterize information (e.g., \textit{truth}, \textit{know}, \textit{hoax}). It is worth noting that despite being an omnipresent pronoun, the word \emph{they} is a highly discriminating word for conspiratorial comments. This denotes the ubiquity of the narrative \emph{they} against \emph{us}. Both \textit{all} and \textit{nothing} in the transcript are also strong indicators for conspiracy, hinting at a lack of nuance. Words that characterize non-conspiratorial content are more random, reflecting the fact that the negative training set is mostly not cohesive.

%
%

%
\begin{figure*}[t]
    \centering
    \includegraphics[width=\linewidth]{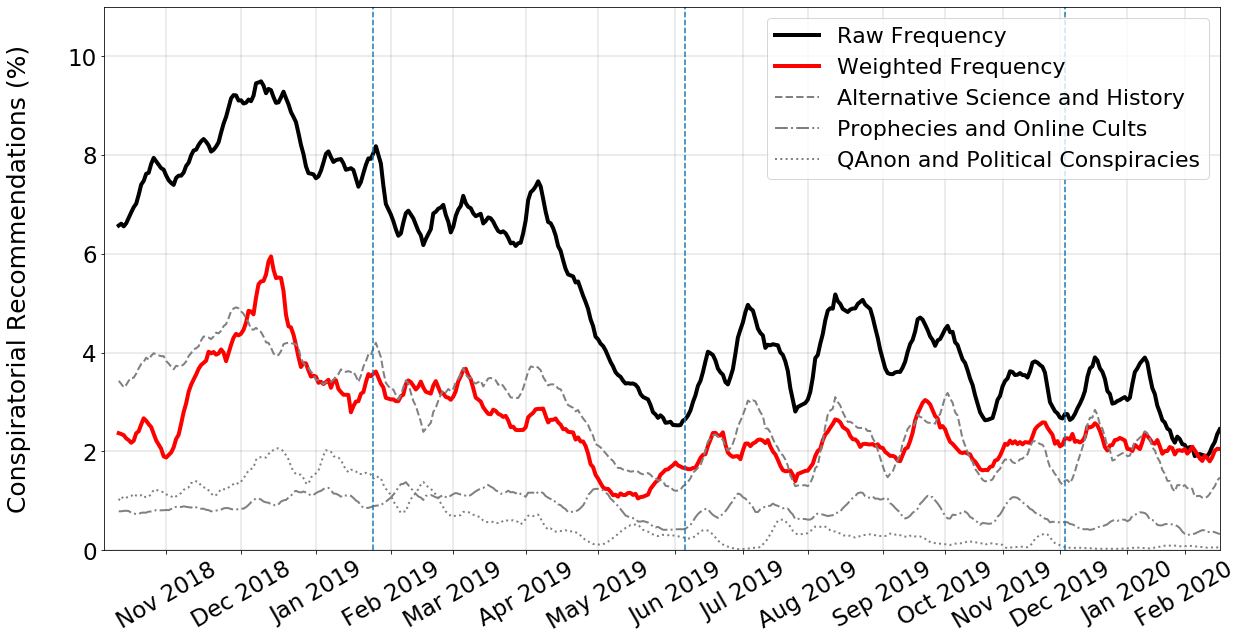}
    \caption{ 
    Longitudinal trends of conspiratorial recommendations from informational channels on YouTube, in which each data point corresponds to a rolling seven day average. The \textbf{raw frequency} is an estimate of the percentage of conspiratorial recommendations obtained by weighting all recommendations classified as conspiratorial by their likelihood. This frequency represents the propensity of the YouTube algorithm to recommend conspiratorial content. The \textbf{weighted frequency} is an estimate of the percentage of conspiratorial weighted by the number of views of the source video. The \textbf{three dashed and dotted lines} correspond to the raw frequency for the top three topics: (1) Alternative Science and History, (2) Prophecies and Online Cults, and (3) QAnon, Deepstate, and New World Order (see Table~\ref{tab:topics}). The \textbf{dotted vertical lines} represent the three YouTube announcements related to their efforts to fight conspiratorial content, on January 25th, June 5th and Dec 3rd of 2019. \newline}
    \label{fig:trend}
\end{figure*}

\begin{table*}
\small
\centering
\begin{tabular}{| p{5 cm} | p{8cm} | p{1.1cm} | p{1.1cm} |}
\hline
\textbf{\hspace{55pt} Topic} & \textbf{\hspace{100pt} Top Words} & \textbf{\% Rec} & \textbf{\% Vid} \\ 
\hline

\textbf{\newline Alternative Science and History} &
\textit{moon, aliens, flat, space, ufo, ancient, nasa, sun, alien, built, pyramids, planet, technology, mars, david, pyramid, water, history, humans, human, science, evidence, energy, sky, stone} &
\textbf{\newline 51.7\%}  &
\textbf{\newline 28.7\%}  \\
\hline

\textbf{\newline Prophecies and Online Cults} &
\textit{jesus, christ, lord, church, bible, shall, spirit, holy, amen, father, pray, satan, heaven, israel, word, brother, son, pastor, temple, unto, rapture, christians, praise, revelation, faith} &
\textbf{\newline 19.1\%}  &
\textbf{\newline 14.9\%}   \\
\hline

\textbf{\newline Political Conspiracies and QAnon} &
\textit{trump, president, wwg1wga, dave, america, country, patriots, bush, deep, mueller, obama, potus, justice, american, law, vote, clinton, hillary, military, fbi, plan, media, democrats, war, gold} &
\textbf{\newline 12.6\%} &
\textbf{\newline 25.9\%}  \\
\hline
\end{tabular}
\vspace*{5mm}
\caption{The top three topics identified by an unsupervised topic modelling algorithm. Each topic is listed with its $25$ most discriminant words and the percentage of recommendations (\% Rec) and videos (\% Vid) that are within this topic among all the conspiratorial videos we identified.}
\label{tab:topics}
\end{table*}

\section*{Results}
\label{sec:results}
We analysed more than $8$ million recommendations from YouTube's watch-next algorithm over $15$ months. Recommendations were collected daily, starting from the most recent videos posted by a set of $1000$+ of the most popular news and informational channels in the U.S. The recommended videos were then fed to a binary classifier trained to detect conspiratorial content based on the video title, description, tags, transcript, and comments (see Methods). The classifier returns the likelihood that a given video is conspiratorial, a score between $0$ (minimal likelihood) and $1$ (maximal likelihood).

\subsection{Longitudinal Trends}

Shown in Fig.~\ref{fig:trend} is our estimate of the percentage of conspiratorial videos recommended by YouTube on information channels, between October 2018 and February 2020 ({\em Raw Frequency}). Also shown is a weighted version of this estimate which accounts for the popularity of the source video ({\em Weighted Frequency}). 

The {\em Raw Frequency} is computed as the product of the number of times a video was recommended and the probability that each video is conspiratorial , Fig.~\ref{fig:precision}. Only videos with a likelihood greater than $0.5$ are counted, providing a conservative estimate (see Methods). The {\em Weighted Frequency} is computed by weighting the {\em Raw Frequency} by the number of views of the source video. This weighting captures the fact that recommendations made from more popular videos have more impact on viewership.

Both of these trends indicate that YouTube experienced a conspiracy boom at the end of 2018. The raw and weighted frequency of conspiratorial recommendations reached a maximum of almost $10\%$ and $6$\%. Fig.~\ref{fig:trend}. Shortly after this, YouTube announced on January 25, 2019 their forthcoming effort to recommend less conspiratorial content.

Starting in April 2019, we monitored a consistent decrease in conspiratorial recommendations until the beginning of June 2019 when the raw frequency briefly hit a low point of $3\%$. Between June and December of 2019, YouTube announced that view-time of conspiratorial recommendations had decreased by $50\%$ and then $70\%$~\cite{blog_2}, a statement mostly consistent with our analysis. The weighted frequency trend that we observed, however, tempers these otherwise encouraging reductions. When the popularity of the source video is accounted for, the proportion of conspiratorial recommendation has steadily rebounded since it's low point in May 2019, and are now only $40\%$ less common than when the YouTube's measures where first announced. 


\subsection{Content}

To understand the nature of the conspiracy videos that we uncovered, we used a topic modelling technique called non-negative matrix factorization (NMF). This algorithm approximates a term-document matrix as the product of a document-topic matrix multiplied by a topic-terms matrix, thus discerning the main topics from the latent semantic structure of the data~\cite{nmf}.

This analysis led to three major topics:  (1) alternative science and history; (2) prophecies and online cults; and (3) political conspiracies. Shown in Table~\ref{tab:topics} are the top $25$ words from the comments section that are the most discriminating to cluster conspiratorial videos in topics (but not to detect conspiracies). 
The first major topic is the redefinition of the mainstream historical narrative of human civilization and development.  This content uses scientific language, without the corresponding methodology, often to reach a conclusion that supports a fringe ideology less well served by facts. For example, the refuting of evolution, the claim that Africa was not the birthplace of the human species or arguments that the pyramids of Giza are evidence of a past high-technology era. Conspiracies relating to climate are also common, ranging from claims of governmental climate engineering -- including chemtrails -- to the idea that climate change is a hoax, and that sustainable development is a scam propagated by the ruling elite. A number of videos address purported NASA secrets, for instance refuting the U.S. moon landing or claiming that the U.S. government is secretly in contact with aliens.

The second topic includes explanations of world events as prophetic, such as claims that the world is coming to an end or that natural catastrophes and political events are religious realisations. Many videos from this category intertwine religious discourse based on scripture interpretations with conspiratorial claims, such as describing world leaders as Satan worshipers, sentient-reptiles or incarnations of the anti-Christ. These videos rally a community around them, strengthened by an \textit{`Us vs. Them'} narrative that is typically hostile to dissenting opinions, in ways similar to cult recruitment tactics~\cite{conspiracy_cults}. We emphasize that most of the religious content found on YouTube does not fall into this category.

The third main topic is comprised of political conspiracies, the most popular of which is QAnon, a conspiracy based on a series of ciphered revelations made on the 4chan anonymous message board by a user claiming to have access to classified U.S. government secrets. These videos are part of a larger set of conspiratorial narratives targeting governmental figures and institutions, such as the Pizzagate, allegations that a deep state cabal and the United Nations are trying to rule a new world order, or claims the Federal Reserve and the media are acting against the interests of the United States.

We found relatively few instances of promotion of conspiratorial videos about the three topics explicitly cited by YouTube in their public statement: flat-earth, miracle cures and 9/11 ~\cite{blog_1}. Other common conspiratorial themes such as alternative theories on the JFK assassination or denial of the Sandy Hook shooting are also rarely promoted. This seems to suggest that highly publicized topics fall under closer scrutiny, while other conspiracies are still regularly recommended.

The three examples listed by YouTube illustrated conspiracies which could \textit{misinform users in harmful ways}. Tribute ought to be paid to YouTube for effectively filtering out some dangerous themes, such as claims that vaccines cause autism. Nonetheless, other themes which we showed to be actively promoted by YouTube were described by the FBI as \textit{very likely to motivate some domestic extremists to commit criminal, sometime violent activity}~\cite{fbi_report}. The report explicitly cites QAnon and Pizzagate conspiracies, depictions of the New World Order, and the United Nations as an organization for the elites to establish a global rule. Similarly, conspiracy-driven online cults have motivated a matricide~\cite{online_cults}. And, seemingly more innocuous conspiracies can also cause unrest, such as when $1.5$ million people gathered on a Facebook group pledging to run onto the military facility Area 51 in a quest to \textit{"see them aliens"}, forcing the U.S. Air Force to threaten them with the use of force~\cite{area51}.

\begin{figure}[t]
    \centering
    \includegraphics[width=\columnwidth]{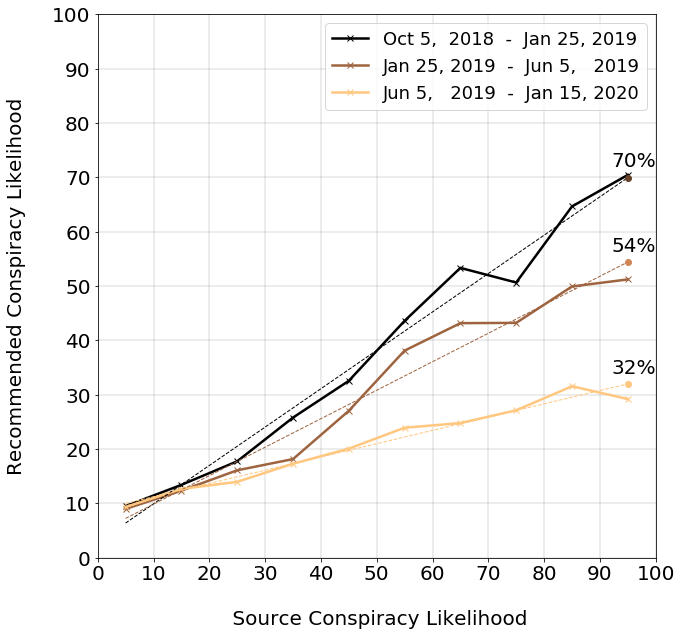}
    \caption{Proportion of conspiratorial recommendations conditioned on the conspiracy likelihood of the source video, for the three time periods between the YouTube announcements. Higher values on the right-side of the graph indicate a stronger 'filter bubble'.}    \label{fig:bubble}
\end{figure}
%
%

\subsection{Filter Bubble}

There is a clear positive correlation between the conspiracy likelihood of the source video and the conspiracy likelihood of the recommended video, Fig.~\ref{fig:bubble}. Although it is an expected feature for a recommendation engine to suggest videos that are similar to the previously watched video, overly selective algorithmic recommendations can lead to a state of informational isolation - a phenomenon called \textit{filter bubble} (or \textit{echo chamber}).

Shown in Fig.~\ref{fig:bubble} is a quantification of this filter-bubble effect in which we see a clear correlation between the proportion of conspiratorial content that is recommended after a conspiratorial video is watched. This correlation is most striking for the time window between October 2018 through January 2019, but has also decreased proportional to the overall reduction shown in Fig.~\ref{fig:trend}.

\section*{Discussion}
\label{sec:discussion}

\subsection{Limitations}

Our data set of recommendations is aimed at emulating the default behavior of YouTube's watch-next algorithm using a set of $1146$ channels as the roots of the recommendation tree. Although this set constitutes a small subset of YouTube and is not necessarily a complete representation of the entire informational landscape, it was obtained through recursive iterations on the watch-next algorithm, starting from the $250$ most followed channels (see Recommendations in Methods). It should, therefore, be by construction, representative of the most commonly recommended informational channels.

Our conspiracy classifier was trained on a set of what we believe to be "clear-cut" conspiracy-theory videos. We recognize that there is an inherent subjectivity in labeling whether a theory is conspiratorial. Many conspiracies, for example, intertwine objective facts with fabricated and far-fetched claims. We have, nevertheless, attempted to categorize a video as conspiratorial based on some objective guidelines (see Data Set in Methods).

Although some topics are more nuanced than others, our conspiracy classifier achieves a relatively high accuracy ($F1 = 0.82$), largely by leveraging the presence of discriminating language as well as references to other conspiratorial topics in the video comments. Nonetheless, the classifier does make mistakes: For instance, videos that debunk conspiracy theories are sometimes mis-classified. We have adjusted for these mistakes by weighting the detected conspiratorial videos by the expected detection accuracy, Fig.~\ref{fig:precision}. Since we have accounted for false positives (incorrectly classifying a video as conspiratorial) but not for false negatives (failing to detect a video as conspiratorial), it is likely that our estimates of conspiratorial content are conservative. Moreover, videos with comments disabled or ones taken down by the author or by the platform before we analyse them are also missing from our analysis, which is more common than average for conspiratorial videos. Another limitation is that personalised recommendations, which we don't account for, can significantly impact the experience for individual users: those with a history of watching conspiratorial content will see higher proportions of recommended conspiracies. Our analysis does not capture weather personalised recommendations have changed at the same pace as non-personalized watch-next recommendations.

\subsection{Policy Implementation}

Our analysis corroborates that YouTube acted upon it's policy and significantly reduced the overall volume of recommended conspiratorial content. The rebound that we observed after the low-point concomitant with the second YouTube announcement~\cite{blog_2} could be caused by (1) Content creators reverse-engineering the moderation process to work around it; (2) YouTube trying to automate a manual moderation process which was in place the previous months; or (3) YouTube relaxing its criteria because of lower engagement or user discontentment. Despite this downtrend over the past year, the overall volume of conspiratorial content recommended from informational channels remains relatively high. Additionally, the weighted frequency of conspiratorial recommendations - which accounts for the popularity of the source video - has rebounded in the months following the YouTube announcement.

Given the massively superior data, labelling and computational resources available to the company, we can affirm that YouTube is technically capable of detecting conspiratorial topics with high accuracy. In fact, for certain topics which seem to fall under particularly close scrutiny, recommended videos are effectively stripped from disinformation. For instance, We did not surface any conspiratorial recommendation about the Corona-virus outbreak, despite the presence of many such videos on the platform. It is encouraging to see that YouTube is willing to target specific issues effectively and in a timely fashion. Deciding what to demote, is therefore a question of policy more than technology.

The reduction of borderline policy is in fact implemented selectively. Our analysis indicates that the selection is mostly happening at the channel level, in line with YouTube's strategy to focus on 'authoritative sources'~\cite{blog_3}. On the one hand, some conspiratorial channels have been completely banned from the default recommendation system. In fact, the ten channels which had most conspiratorial recommendations before April 2019~\footnote{Anonymous Official, X22Report, Disclosed TruthTV, Edge of Wonder, Truthstream Media, ZEG TV HIDDEN FROM THE PUBLIC, Matrix Wisdom, THAT IS IMPOSSIBLE, David Icke, UAMN TV} and together accounted for more than $20\%$ of all recommended conspiracies now make up for less than $0.5\%$ recommended conspiracies.

On the other hand, since the policy update a set of five channels~\footnote{Fox News, Science Channel, London Real, The Nimitz Encounters, After Skool} account for $25\%$ of all conspiratorial recommendations, whereas they previously represented less than $0.5\%$. Most of these rising channels intertwine legitimate videos and conspiracies, and seem to be benefiting from a white-listed status to have increasingly borderline content be recommended. Many of these large channels can funnel traffic from mainstream channels, which explains why the gap between the raw and the weighted frequency has been narrowing, Fig.~\ref{fig:trend}. Lastly, some conspiratorial channels have been continuously recommended throughout our analysis and seem to have escaped notice, including some that promote particularly insidious disinformation.\footnote{Perry Stone, A Rood Awakening!, Sid Roth's It's Supernatural!, Zohar StarGate Ancient Discoveries, DTBM OnlineVideoTraining}

\subsection{Filter Bubble}

It is true that YouTube, overall, no longer recommends conspiratorial videos with a higher likelihood than what was previously watched~\cite{mohan}. This result is in line with recent research which did not find strong quantitative evidence of a systematic push towards right-wing political content~\cite{pathways}. Our analysis, however, shows that after a conspiratorial video is clicked, there is a high - yet decreasing - likelihood that the algorithm will recommend another one. For those with a history of watching conspiratorial content, the filter-bubble effect is strongly reinforced by personalized recommendations, which we don't capture in this study~\cite{Zhao2019}.

It is hard to say if this selective exposure is more pronounced for conspiratorial content than for other categories. Filter bubbles and its impact in shaping political opinions might have been overstated~\cite{selective_exposure,supply_demand}, but the fact that the filter-bubble effect~(Fig.~\ref{fig:bubble}) has decreased over the past year in proportions similar to the raw frequency indicates that it might have been an important driver of conspiratorial viewership. Moreover, we argue that the repercussions of selective exposure may be stronger with conspiracy theories than they are with more typical political content, because conspiratorial narratives are rarely challenged or even addressed on other media. Conspiracy theories also tend to be unfalsifiable in the sense that evidence against a conspiracy can often be interpreted as evidence of its truth. Presenting opposing views, therefore, may not affect the faith in the conspiracy. 

\subsection{Summary}

The overall reduction of conspiratorial recommendations is an encouraging trend. Nonetheless, this reduction does not make the problem of radicalization on YouTube obsolete nor fictional, as some have claimed~\cite{ledwich}. Aggregate data hide very different realities for individuals, and although radicalization is a serious issue, it is only relevant for a fraction of the users. Those with a history of watching conspiratorial content can certainly still experience YouTube as filter-bubble, reinforced by personalized recommendations and channel subscriptions. In general, radicalization is a more complex problem than what an analysis of default recommendations can scope, for it involves the unique mindset and viewing patterns of a user interacting over time with an opaque multi-layer neural network tasked to pick personalized suggestions from a dynamic and virtually infinite pool of ideas.

With two billion monthly active users on YouTube, the design of the recommendation algorithm has more impact on the flow of information than the editorial boards of traditional media. The role of this engine is made even more crucial in the light of (1) The increasing use of YouTube as a primary source of information, particularly among the youth~\cite{pew}; (2) The nearly monopolistic position of YouTube on its market; and (3) The ever-growing weaponization of YouTube to spread disinformation and partisan content around the world~\cite{oii}. And yet, the decisions made by the recommendation engine are largely unsupervised and opaque to the public. 

This research is an effort to make the behavior of the algorithm more transparent so that YouTube can be held accountable for their statements.~\footnote{The full list of recommended videos we detected with a conspiracy likelihood above 0.5 is available at \mbox{\url{https://github.com/youtube-dataset/conspiracy}}} We hope it will fuel a public discussion, not about whether YouTube should allow for conspiratorial content on the platform, but about whether such content is appropriate to be part of the baseline recommendations on the informational YouTube

\bibliographystyle{SageV}
\bibliography{conspiracy}

\end{document}